\documentclass[useAMS,usenatbib]{mn2e}
\usepackage{graphicx}
\usepackage{epstopdf}
\usepackage{color}

\title[Recent star formation in BCGs]{Evidence for recent star formation in BCGs: a correspondence between blue cores
and UV excess}

\author[Pipino et al.]{A. Pipino\thanks{current address: pipino@usc.edu}$^{1,2}$, S. Kaviraj$^1$, C. Bildfell$^3$,
A. Babul$^3$, H. Hoekstra$^{3,4}$\thanks{Alfred P. Sloan fellow},
 \& J. Silk$^1$ \\
$^1$Astrophysics, University of Oxford, Denys Wilkinson Building,
    Keble Road, Oxford, OX1 3RH, U.K.\\
$^2$Department of Physics \& Astronomy, University of Southern California, Los Angeles 90089-0740, USA\\
$^3$Department of Physics \& Astronomy, University of Victoria, Victoria, BC V8P 1A1, Canada\\
$^4$Sterrewacht Leiden, Leiden University, Niels Bohrweg 2, NL-2333 CA Leiden, the Netherlands}

\long\def\symbolfootnote[#1]#2{\begingroup%
\def\thefootnote{\fnsymbol{footnote}}\footnote[#1]{#2}\endgroup}

\begin{document}

\date{Accepted 2009 January 20.  Received 2008 December 20; in original form 2008 July 17}

\pagerange{\pageref{firstpage}--\pageref{lastpage}} \pubyear{2008}

\maketitle

\label{firstpage}

\begin{abstract}
We present a joint analysis of near-ultraviolet (NUV) data from the
GALEX mission and (optical) colour profiles for { a sample of 7
Brightest Cluster Galaxies} (BCGs) in the Canadian Cluster
Comparison Project. We find that every BCG which has a blue
rest-frame UV colour also shows a blue-core in its optical colour
profile. Conversely, BCGs that lack blue cores and show monotonic
colour gradients typical of old elliptical galaxies, are red in the UV.  We
interpret this as evidence that the NUV enhancement in the blue
BCGs is driven by \emph{recent star formation} and not from old
evolved stellar populations such as horizontal branch stars.
Furthermore, the UV enhancement cannot be from an AGN because the
spatial extent of the blue cores is significantly larger { than the
possible contamination region due to a massive black hole}. The recent
star formation in the blue BCGs typically has an age less than 200 Myrs
and contributes mass fractions of less than a percent.
Although the sample studied here is small, we demonstrate,
\emph{for the first time}, a one-to-one correspondence between
blue cores in elliptical galaxies (in particular BCGs) and a NUV-enhancement observed
using GALEX.   The combination of this one-to-one correspondence and  the consistently young
age of recent star formation, coupled with additional correlations with the host cluster's X-ray properties, 
strongly suggests that the star formation is fueled by gas cooling out of the intracluster medium.
In turn, this implies that any AGN heating of the intracluster medium in massive clusters 
only acts to reduce the magnitude of the cooling flow and that once this flow starts, it is
nearly always active.  Collectively, these results suggest that AGN feedback in present-day
BCGs, while important, cannot be as efficient as suggested by  the recent theoretical model by proposed by 
De Lucia et al. (2006).
\end{abstract}

\begin{keywords}
galaxies: clusters: general -- galaxies: elliptical and lenticular, cD -- galaxies: evolution -- cooling flows -- X-rays: galaxies: clusters
\end{keywords}

\section{Introduction}

Elliptical galaxies provide a critical
test bench for many scenarios of galaxy formation because they
follow tight relations in their photometric (optical), chemical and
dynamical properties.  Within massive groups and clusters, elliptical 
galaxies populate a tight region of the colour-magnitude relation (CMR) 
(Bower et al. 1992) known as the red sequence, which is mainly driven 
by an underlying `mass-metallicity' relation (e.g. Carollo et al. 1993).
The lack of significant change in the slope and scatter of the CMR 
(e.g. Stanford et al., 1998), the slow evolution of colours (e.g. Saglia et al. 
2000; Ellis et al., 1997),  and line strength indices of cluster early-type
galaxies out to $z\sim 1$ (e.g. Bernardi et al., 2003) indicate
that the \emph{bulk} of their stars that comprise these systems 
formed over a relatively short period of time at high redshifts 
($z>3$) (see Renzini, 2006 for a recent review), though the issue of where 
these stars formed and how they  were assembled into the observed  elliptical 
galaxies has yet to be  conclusively resolved.

The simplest model for the formation of elliptical galaxies  --- especially the most 
massive ones ---  is the `monolithic collapse' model (see e.g. Larson 1974,  Matteucci 
\& Tornambe 1987, Pipino \& Matteucci 2004, Merlin \& Chiosi, 2006).  In 
this model, the elliptical galaxies are believed to have formed from homogeneous 
collapse of intergalactic material, followed by a rapid, massive starburst, during 
which nearly all of the available gas is converted into stars that comprise the 
observed stellar content of the elliptical galaxies today.   In its simplest form, the 
monolithic model predicts that the vast majority of the 
present-day elliptical systems were already be in place by $z\sim 3$ and that once 
formed, have been evolving passively towards the present.  Consequently, this model 
predicts that ellipticals galaxies should be largely \emph{red and dead} today.  

Detailed recent observations of elliptical galaxies, however, indicate a more complex
formation history:   Taylor et al. (2008), for example, find that at the most only 20\% 
of the local red sequence galaxies with stellar mass  $M_* > 10^{11}\;M_\odot$ were 
already in place by $z\sim 2$  (but see Whiley et al. 2008, who do not detect any 
significant change in the stellar mass of the BCG since $z\sim 1$).   Kormendy et al. 
(2008) argue that the mass-dependent  variations in the structural properties of elliptical 
galaxies is inconsistent with the idea  of a uniform, synchronized formation history.  And 
using rest-frame  UV observations\footnote{In contrast to the optical spectral 
ranges, the UV is highly  sensitive to even  small fractions of young stars 
(younger than about a Gyr), making it an excellent probe of the low-level 
recent star formation.} to  probe the finer details of the  star formation histories  in 
elliptical galaxies, Kaviraj et al. (2007a; 2008) find compelling evidence that 
while the bulk of the stars that  comprise the elliptical galaxies are old, the galaxies continue
to form stars at a reduced rate over the lifetime of the Universe.  Specifically, using 
GALEX (UV) and SDSS (optical) photometry, Kaviraj et al. (2007a) find that \emph{at least} 
30\% of nearby ($0<z<0.11$),  massive  ($M_{\rm r}<-21$) ellipticals  show \emph{unambiguous} 
signatures of star formation within the last Gyr, contributing up to a few percent of the 
stellar mass of the galaxy. 

An alternative model for the formation and evolution of galaxies is the hierarchical 
clustering model in which the these systems are \emph{assembled} via successive mergers
of lower mass systems, with both the formation of the lower mass progenitors occurring at 
an earlier epoch and the merger rate being larger in regions of the Universe characterized by 
higher than average total mass density.  The hierarchical model is an inescapable prediction of the currently 
favored cold dark matter models for the formation and evolution of large-scale structure in the 
Universe.  Strictly speaking, the hierarchical model only speaks to how the mass is assembled, not
to the details of star formation, which is what the observations are sensitive to.  The 
latter depends on a series of complex, highly nonlinear baryonic physics, including the manner in which
galaxies acquire their gas (Keres et al. 2005), the local environment where the galaxy and its progenitors 
reside, the details of the cooling and heating processes that the gas is subjected to (e.g.~Silk 1977;  White 
\& Rees, 1978; Scannapieco et al. 2005;  Kaviraj et al. 2005, 2007b), etc.   The most recent detailed 
study of the formation history of elliptical galaxies by De Lucia et al. (2006) shows 
that  while the bulk of the  stars that make up the most massive ellipticals form rapidly and high 
redshifts, they do so in a number of different progenitor systems and typically, the assembly of the 
progenitors into a single massive elliptical galaxy occurs much later.  Whether the observed  structural, 
chemical and photometric properties of massive  elliptical galaxies  is compatible with hierarchical 
late-time assembly remains to be seen  (see, for example, Cimatti et al. 2006;   Ciotti et al. 2007; 
Pipino \& Matteucci 2008; Pipino et al. 2009, Maiolino et al. 2008, and Whiley et al. 2008 for observational arguments 
against late-time assembly).

In the context of the hierarchical models, the most important of the recent innovations is the inclusion
of AGN feedback.  AGN feedback plays a central role in driving gas out of the progenitor systems,
thereby both truncating the initial starburst in these systems and ensuring that any subsequent mergers
involving these objects are gas-poor (i.e.~"dry mergers").  AGN feedback is also critical for preventing 
the cooling of any gas that subsequently accumulates in the halos of the massive elliptical galaxies or the larger 
systems (e.g.~groups and clusters) that they are embedded in.  Prior to the inclusion of AGN feedback, the 
hierarchical model produced elliptical galaxies that were both  much too luminous and much too blue 
compared to the present-day population of massive elliptical galaxies.    Unfortunately, details of how AGN 
feedback actually works is not well understood and the properties of the elliptical galaxies in the hierarchical
model depends sensitively on how AGN feedback is modeled.

In this paper, we will use recent observations of a  special class of massive elliptical galaxies, the Brightest Cluster 
Galaxies (BCGs), to  gain additional insights into processes shaping the formation of elliptical galaxies.
We are especially interested knowing whether these systems show evidence of late-time star formation, the 
conditions that facilitate late-time star formation and in understanding the nature of AGN feedback in these
systems, at least at the present-time. As the most massive, most luminous galaxies in their local environment,\footnote{It is often 
incorrectly assumed that BCGs always reside at the center of the cluster potential.   As shown by Bildfell et al. 
(2008), this is not necessarily so.}  these galaxies have long been of special interest.   In spite of their special 
status, they have much in common with other massive early-type 
galaxies.  For example, the mean stellar ages and metallicities of BCG galaxies are similar to those of non-BCG
ellipticals of the same mass (Fisher et al., 2005; Brough et al. 2007, von der Linden et al., 2007).   More importantly,
within the context of the hierarchical model, the essential features of the formation history of the BCGs is predicted to
be similar to those of  massive non-BCG early-type systems.  Specifically, while nearly 80\% of the stars that 
comprise the BCGs form at $z > 3$, they form in several distinct systems.   The BCGs themselves do not take on a
distinct identity  until after $z\sim 0.7$. (De Lucia \& Blaizot 2007; Romeo et al. 2008).   

{
The BCGs that emerge from the theoretical modeling by De Lucia \& Blaizot (2007) are, for all intensive purposes,
\emph{red and dead}.  They experience virtually no star formation after the initial burst because the modeling assumes
an extremely efficient form of AGN feedback.  However, several recent studies (non-UV based) have reported
examples of ongoing star formation in the brightest cluster galaxies (Cardiel et al. 1998, Crawford et al. 1999, Edge 2001, 
Goto 2005, McNamara et al. 2006, Wilman et al. 2006, { O'Dea et al. 2008}, Bildfell et al. 2008, Cavagnolo et al. 2008, 
Rafferty et al. 2008); most of these BCGs reside in cool core
clusters.  { Additionally,  Hicks \& Mushotzky (2005) noted an excess in the UV flux ---
as determined from the XMM-Newton Optical Monitor --- in many (but not all) the cooling flow clusters in their sample,
which they interpreted as evidence for star formation.}   Most recently,  Bildfell et al. (2008) undertook a comprehensive 
study of 48 clusters that span a wide range of X-ray characteristics.  Specifically, they analyzed the surface brightness and
color profiles of the BCGs hosted by these clusters, seeking to relate the resulting trends to the relative location of the BCGs within 
the clusters as well as to their global X-ray luminosity ($L_x$) and temperature ($T_x$).  They
found that 25\% of their BCGs had bluer colors in their central regions, which they interpreted as evidence for ongoing
star formation.   They also found that these blue core systems only occurred in cool core clusters (see also Cavagnolo et al 2008)
and then only if the BCG is located at the cluster center.  For completeness, we note that cool core clusters  are systems whose 
central gas density is sufficiently high that the corresponding  radiative cooling timescale is $< 5\;$Gyrs. (See  McCarthy et al. 
2004; 2008 for a detailed discussion of the diversity in the cluster population, including the distinction between 
cool core versus non-cool core clusters).

The mounting evidence for active star formation poses a challenge for models that invoke strong AGN feedback.   Bildfell
et al (2008) suggest that in the systems that they studied, heating by AGN feedback may be offsetting most of the radiative
losses suffered by the hot gas surrounding the BCGs but not all.  Therefore, the gas cools but at a significantly reduced 
rate.  If this is indeed the case, it represents an important clue into how AGN feedback operates.   Our long-term goal, 
therefore, is to understand that implications of ongoing star formation in BCGs within the broader context of galaxy 
formation.   As a first step, we need to establish that the blue core phenomenon is indeed associated 
with star formation.   As discussed below, we do so by demonstrating that the blue core are unambiguously linked 
to UV-enhancement.   Next we quantify the extent of recent star formation.
Specifically, we are interested in determining the age of the last star formation event and the fraction of the total stellar 
mass that it gave rise to.   Finally, we review the conditions characterizing the environments of the BCGs in our
sample and use the apparent correlations to identify the source(s) of the cold gas that is fueling the residual star 
formation.

In order to carry out our analysis, we start by cross-matching the 48 BCGs in the Bildfell
et al. catalog with archival data from the GALEX GR3. We find 10
BCGs in common, of which we analyze 7.   In Sec.~\ref{data}, we briefly summarize the 
main characteristics of the Bildfell et al. sample and the data
retrieved from the GALEX archive.   In Sec.~\ref{model}, we present the model set up used to estimate
the characteristics of the recent star formation (ages and mass
fraction) in each BCG. Finally, our results are discussed in
Sec.~\ref{results}, and the conclusions are presented in
Sec.~\ref{conclusions}. The cosmological parameters used
throughout this work are $H_0=70$ km s$^{-1}$ Mpc$^{-1}$,
$\Omega_m=0.3$ and $\Omega_{\Lambda}=0.7$.


\section{Data}
\label{data}

The galaxies studied by Bildfell et al. (2008) are drawn from the
Canadian Cluster Comparison Project (CCCP) The CCCP is a study of
an X-ray selected sample of 50 clusters from the Advanced Satellite for Cosmology  and Astrophysics (ASCA)
catalog of Horner (2001),
with redshifts in the range $0.15 \leq z \leq 0.55$ and most
clusters having X-ray temperatures $T_x
> 5$ keV. Almost all clusters have also either Chandra
or XMM-Newton high resolution data.
The CCCP clusters span the observed
range in scatter of cluster X-ray and S-Z scaling relations, and thus create maximal leverage for the
investigation of baryonic feedback effects. The combination of
X-ray and deep, multi-filter optical data makes the CCCP an
excellent sample for studying the links between BCGs and their
host cluster environments.  The deep optical imaging data was
obtained using the Canada-France-Hawaii Telescope (CFHT). The full
sample comprises a set of 30 clusters observed in $g'$ and $r'$
filters using the MegaCam detector and 20 clusters observed in the
$B$ and $R$ filters using the CFH12K detector which were taken
from the CFHT data archive at the Canadian Astronomical Data
Centre (CADC).

A quarter ($13/53$) of the BCGs in the CCCP show colour profiles that
become increasingly blue in the core. These galaxies are displaced
from their host-cluster optical red sequence by $\sim 0.5$ to
$1.0$ $mag$ in ($g'$-$r'$), which is important for optical cluster
studies that may reject the BCGs based on colour.  The blue cores
in these systems cannot be explained by AGN point-sources because
(a) they are at least twice the size of the seeing disk and (b)
Type II AGN do not contaminate the optical spectrum  at more than
a few percent level  (Zakamska et al. 2006a; 2006b).
The above arguments also apply to the NUV, in that 
the contamination due to AGN is less than 15\% in the UV (see Salim et al. 2007 for
a more detailed discussion).  Moreover, Kauffmann et al. (2007)
examined a sample of nearby AGN hosts and find that their UV emission too
is quite extended, thus unlikely to emanate from the central 
AGN, whereas it seems to be associated with star forming discs. For these reasons, we shall assume that all of the 
NUV flux is purely stellar in origin and compute star formation rates accordingly.
Strictly speaking, however, the star formation rates that we derive represent 
upper limits.

\begin{table*}
\centering
\begin{minipage}{175mm}
\caption{Galaxies from Bildfell et al. (2008) present in GALEX DR2}
\begin{tabular}{|l|l|l|l|l|l|l|l|l|l|c|c|}
\hline
BCG & Name & $z_{sp}$ & $\alpha_{BCG}$ & $\delta_{BCG}$ & $r'$,$R$ & $g'$,$B$ & $r_e^{r'}$ &  Core & $R_*$ & NUV ($2300\AA$) & exp \\
ref& & & (J2000) & (J2000) & mag & mag & kpc & colour & kpc & mag & ks \\
\hline
\hline
1&Abell 1835 & 0.25 & 14 01 02.10 & $+02$ 52 42.69 & $15.97$ & 17.17 & $66.4^{+0.6}_{-0.1}$ &  blue & $19\pm2$ & 19.13 $\pm$ 0.02 & 1.7\\
2&Abell 1942 & 0.22 & 14 38 21.88 & $+03$ 40 13.34 & $16.35$  & 17.38 & $43.7^{+0.2}_{-0.2}$ &  red & - &
22.68 $\pm$ 0.28 &  1.7 \\
3&Abell 2111 & 0.23 & 15 39 40.52 & $+34$ 25 27.46 & $ 17.26$ & 18.50 &  $26.4^{+0.3}_{-0.3}$ &  red  & - &
23.41 $\pm$ 0.35 &  7.3 \\
4&CL0910+41 & 0.44 & 09 13 45.52 & $+40$ 56 28.54 &  $ 18.93$ & 18.81 & $32.6^{+0.5}_{-0.4}$ &   blue & $40\pm8$ & 19.58 $\pm$ 0.11 & 0.4 \\
5*&MS0440+02 & 0.19 & 04 43 09.92 & $+02$ 10 19.33 & $ 15.06$ & - & $102.0^{+0.3}_{-0.3}$ &  red & - &
21.84 $\pm$ 0.44 & 0.09 \\
6*&MS0451-03 & 0.54 & 04 54 10.84 & $-03$ 00 51.39 & $ 19.38$ & 20.75 & $45.5^{+0.3}_{-0.6}$ & blue & $20\pm2$ & 20.65 $\pm$ 0.28 & 0.11 \\
\hline
7&Abell 2219 & 0.2256 & 16 40 19.85 & $+46$ 42 41.30 & $15.98$ & 18.40 & $61.7^{+0.5}_{-0.5}$ &  red & - &  22.66 $\pm$ 0.27 & 2.75 \\
8&Abell 2390 & 0.2280 & 21 53 36.84 & $+17$ 41 44.10 & $17.33$ & 19.22 & $18.4^{+0.3}_{-0.2}$ & blue & $25\pm5$ &20.94 $\pm$ 0.07 & 1.96 \\
9*&CL0024+16 & 0.39 & 00 26 35.68 & $+17$ 09 43.48 &   $17.45$ & - & $60.1^{+1.8}_{-0.8}$ & red & - &
21.91 $\pm$ 0.09 & 15.42 \\
10&MS1512+36 & 0.3727 & 15 14 22.51 & $+36$ 36 21.30 & $17.66$ & 19.94 & $57.6^{+0.9}_{-0.9}$ & blue & $15\pm2$ &21.40 $\pm$ 0.34  & 0.07 \\
\hline
\end{tabular}

{\small CCCP targets from Bildfell et al. (2008). The first group
of six BCGs was observed with CFHT MegaCam, while the second group
was observed with the CFH12K camera. The galaxies marked with a * 
have not been used for the analysis (see Section 2). 
Redshifts ($z$) listed are spectroscopic. The ($r'$, $g'$) and ($R$, $B$) apparent magnitudes
are in AB and Vega system, respectively (see text). The associated
errors are typically 0.03 mag.  Note that for the reasons outlined in section \ref{data} no $g'$ or $B$ magnitudes 
were obtained for CL0024+16 and MS0440+02. 
The $r'$ best-fit surface brightness profile
parameter $r_e^{r'}$ is given to be compared to the size of the blue region 
$R_*$ - if a blue core is present. Core colour is given as `red' or `blue' based on
the shape of the inner colour profile (see Bildfell et al.,
2008). Galex NUV magnitudes are in AB system.
Galex NUV exposure times (exp) in units of 1000s.} \label{tab:MegaCamdata}
\end{minipage}
\end{table*}
\normalsize

The Bildfell et al. sample was cross-matched with publicly
available $UV$ photometry from the second data release of the
GALEX mission (Martin et al. 2005). GALEX provides two $UV$
filters: the far-ultraviolet ($FUV$), centered at $\sim1530\AA$ and
the near-ultraviolet ($NUV$), centered at $\sim2310\AA$. Note that,
since the bulk of our sample are at intermediate redshifts
($z>0.15$) we only use the NUV filter in this study, since the NUV
filter traces the spectrum between rest-frame NUV and FUV. The
cross-matching produced 10 BCGs which have \emph{at least} a
detection in the $NUV$ filter. { The remainder of the galaxies are
not observed partly because of a lack of sufficient spatial coverage in Galex data and, possibly, short exposure times in some Galex fields.} The positional matching is
performed within the GALEX fiducial angular resolution of 6". NUV
magnitudes and exposure times for the GALEX fields are given in
Table \ref{tab:MegaCamdata}, along with other information related
to the galaxies such as redshifts, J2000 coordinates, effective
radii, and the type of core (red or blue) as described in Bildfell
et al. In each case the GALEX image with the longest exposure time
was used.

We note that, in general, galaxies harbouring recent star formation are more likely
to be detected than non recent star formation-galaxies (since their UV flux will be
higher), especially at higher redshift. Thus at any redshift the
chances of detecting a UV blue BCG is higher than its UV red
counterparts. However, since the goal of this work is to verify if
a strong $NUV$ flux is unambiguously linked to blue-core galaxies
(and vice versa) the incompleteness in the detection of UV red
BCGs does not affect our results.

Out of the 10 BCGs that are detected by GALEX, 5 have blue cores
and 5 have normal (red) cores. However, this sample reduces to 7
objects because we have to exclude the BCGs in MS0440, CL0024 and
MS0451 for the reasons outlined below.
The MS0440 and CL0024 BCGs
are excluded because there are systematics due to crowding in both
of these systems. Moreover, the MS0440 BCG lies right at the edge
of a very bright stellar reflection artifact, which causes
problems for proper background subtraction.

In the case of the MS0451, there is a large foreground
spiral that overlaps with its BCG. To minimize contamination,
Bildfell et al. were required to mask a significant fraction of
the BCG before measuring a total magnitude from its optical image.
The central regions of the BCG are more heavily masked than the
outer regions, which leads the colour to be dominated by that of
the outer regions making the galaxy optically red. Since we
extract the UV photometry directly from the GALEX GR3 database
(without any masking) and given the large PSF of GALEX ($\sim$ 6
arc seconds) it is likely that the star formation in the
overlapping spiral contributes to the UV signal we measure. This
results in the MS0451 BCG having a very red ($g'-r'$) colour and an
extremely blue UV-optical colour, rendering it difficult to
perform accurate parameter estimation using realistic models.


\section{Modelling the spectrum: quantifying the age and mass fraction in young stars}
\label{model} 

We estimate parameters governing the star formation history (SFH)
of each galaxy by comparing its GALEX $NUV$ and CFHT optical
photometry to a library of synthetic photometry, generated using a
large collection of model star formation histories.  
Specifically, we follow Kaviraj et al. (2007b) and make use of 
$g'$-$r'$ \& ($NUV$-$r$) colours, with the near-UV constraining the amount of
recent star formation and the red optical light providing the normalization.

Each model SFH
is constructed by assuming that a burst of star formation at high
redshift ($z\sim3$) is followed by a second burst, which is
allowed to vary in age between 0.001 Gyrs and the look-back time
corresponding to $z=3$ in the rest-frame of the galaxy, and mass
fraction between 0 and 1. Both bursts are assumed to be
instantaneous.  A Salpeter (1955) initial mass function is adopted for
each burst.  Figure \ref{fig:sfh_cartoon} shows a schematic
representation of the model SFHs.  { As an alternative, we could have 
modeled the second star formation  event  as extended in time (or 
even, continuous).   We are unable to do so, however,  because this 
approach requires more wavelength coverage than  currently available 
in order to break the  degeneracies between age, metallicity and dust content.}

\begin{figure}
\begin{center}
\includegraphics[width=3in]{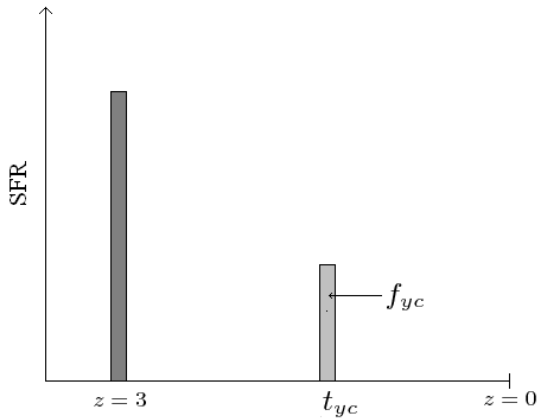}
\caption{Model SFHs (see Section 3) are constructed by assuming
that an instantaneous burst of star formation at high redshift
($z=3$) is followed by a second instantaneous burst which is
allowed to vary in age and mass fraction. The main free parameters
are the age ($t_{yc}$), mass fraction ($f_{yc}$) of the second
burst. $t_{yc}$ is allowed to vary from 0.001 Gyrs to the
look-back time corresponding to $z=3$ in the rest-frame of each
galaxy. $f_{yc}$ varies between 0 and 1.}\label{fig:sfh_cartoon}
\end{center}
\end{figure}

\begin{figure}
\begin{center}
\includegraphics[width=3in]{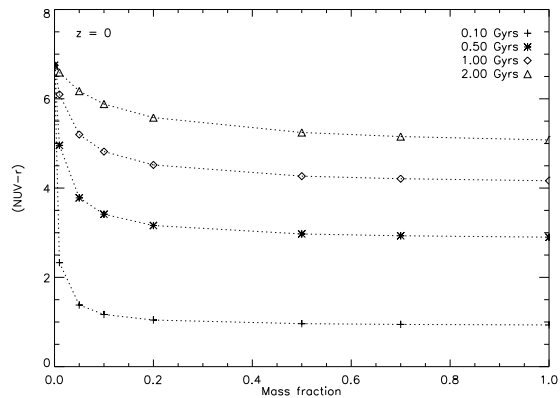}
\caption{The sensitivity of the UV to young stars. We assume two
instantaneous bursts of star formation, where the first burst is
fixed at $z=3$ and the second burst is allowed to vary in age and
mass fraction. The near-UV (NUV) colour of the composite stellar
population is plotted as a function of the age (symbol type) and
mass fraction (x-axis) of the second burst. It is apparent that
even a small mass fraction ($\sim$1\%) of young stars ($\sim0.1$
Gyrs old) causes a dramatic change in the $NUV-r$ colour compared
to what might be expected from a purely old stellar population
($NUV-r\sim 6.8$).} \label{fig:uv_sense}
\end{center}
\end{figure}

To build the library of synthetic photometry, each model SFH is
combined with a single metallicity in the range 0.1Z$_{\odot}$ to
2.5Z$_{\odot}$ and a value of dust extinction parametrised by
$E_{B-V}$ in the range 0 to 0.5. Photometric predictions are
generated by combining each model SFH with the chosen metallicity
and $E_{B-V}$ values and convolving with the stellar models of Yi
(2003) through the GALEX $NUV$ and CFHT optical filters. 
The model library contains $\sim750,000$ individual models.
Finally, since our galaxy sample spans a range of redshifts,
equivalent libraries are constructed at the redshift of every
galaxy.

The primary free parameters in this analysis are the age
($t_{yc}$) and mass fraction ($f_{yc}$) of the second burst, which
provides the young component (`yc') of the stellar population in
the galaxy.  For each galaxy, parameters are estimated by comparing
each observed galaxy to every model in the synthetic library, with
the likelihood of each model ($\exp -\chi^2/2$) calculated using
the value of $\chi^2$ computed in the standard way. From the joint
probability distribution, we marginalise over the dust and
metallicity to extract the $t_{yc}$ vs. $f_{yc}$ probability space
(see next section) for each galaxy. This space shows, for every
galaxy, the probability that the recent star formation is described by a particular
value of $t_{yc}$ and $f_{yc}$. This allows us to explore the
characteristics of the recent star formation in the individual
BCGs and correlate them to the presence (or absence) of a blue
core in the optical image.
The error in the $\chi^2$ is calculated by adding, in quadrature, 
the observational uncertainties and typical errors in the stellar models, 
which we assume to be 0.05 mag in each optical filter and 0.1 mag for the GALEX UV filter (Yi 2003).
 Note that the method is similar in
design to past techniques designed to detect small `frostings' of
young stars over a underlying old stellar population, either using
rest-frame UV data (e.g. Ferreras and Silk 2000) or spectroscopic
line indices (e.g. Trager et al. 2000).

The sensitivity of the UV to young stars is demonstrated in Figure
\ref{fig:uv_sense}. We assume two instantaneous bursts of star
formation, where the first burst is fixed at $z=3$ and the second
burst is allowed to vary in age and mass fraction. The near-UV
 colour of the composite stellar population is
plotted as a function of the age (symbol type) and mass fraction
(x-axis) of the second burst. It is apparent that even a very
small mass fraction ($\sim$1\%) of young stars ($\sim0.1$ Gyrs
old) causes a dramatic change in the $NUV-r$ colour compared to
what might be expected from a purely old stellar population
($NUV-r\sim 6.8$) or from the analysis of any optical
colour-magnitude relation.
Given that typical observational uncertainties
in the $NUV-r$ colours from modern instrumentation are $\sim0.2$
mag, the usefulness of the UV in detecting residual amounts of recent star formation
becomes quite apparent. Note that the $NUV$ filter is taken from
the GALEX filter set.

\section{Results and discussion}
\label{results}

In Figure \ref{fig:colours}, we present the UV-optical
colours of the 7 BCGs in our final sample. In Figures
\ref{fig:prob-red} and \ref{fig:prob-blue}, we present the
quantitative SFH parameter estimation in terms of $t_{yc}$ and
$f_{yc}$ for the red-core and blue-core BCGs respectively.\footnote{  
We caution the reader that the 
estimates of $t_{yc}$ and $f_{yc}$ for the BCG in MS1512 are not particularly 
well constrained because of the large error in NUV flux due to extremely short 
exposure --- see Table 1.}

Since the BCGs span a large range in redshift we do not simply
show their  ($NUV-r$) colours because their `red
sequence' positions (i.e. the position of a old, passively
evolving population) are very different. Instead, we show, in
Fig.~\ref{fig:colours}, the \emph{difference} between the
 ($NUV-r$) colour of each galaxy and the `red sequence'
position at its given redshift, calculated using a dustless,
solar-metallicity simple stellar population (SSP) forming at
$z=3$. 
The local early-type population (Kaviraj et al. 2007a) is shown over-plotted using the small
black dots. The value of $M_{\rm r}$ for the BCGs presented here is
calculated by converting from the magnitudes in the CFHT Mould and
Megaprime $r$-band filters (from Bildfell et al.) using a
solar-metallicity SSP that forms at $z=3$. Note that this local
population is drawn mainly from the field and its local density
distribution varies strongly from that of the BCG population (see
Figure 5 in Schawinski et al. 2007).

\begin{figure}
   \centering
   \includegraphics[width=3.5in]{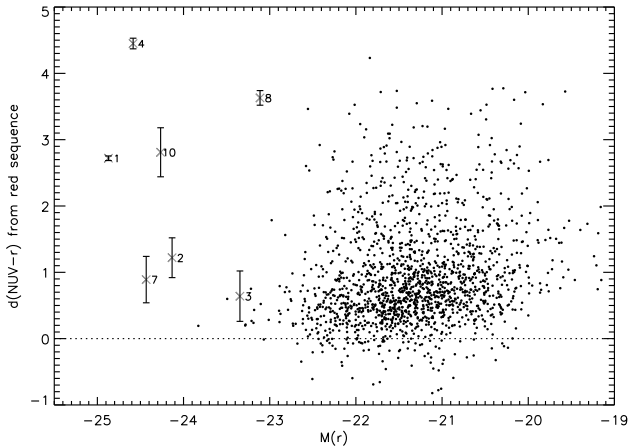}
   \caption{The difference between  the red sequence
position  of each BCG at its given redshift, calculated using a solar
metallicity simple stellar population (SSP) which forms at $z=3$, and the
 ($NUV-r$) colour.
The local early-type population is shown over-plotted using the small lack
dots. The numbers identify the galaxies according to the order in
Table 1. }
   \label{fig:colours}
\end{figure}
\begin{figure}
$\begin{array}{c}
   \includegraphics[width=3.5in]{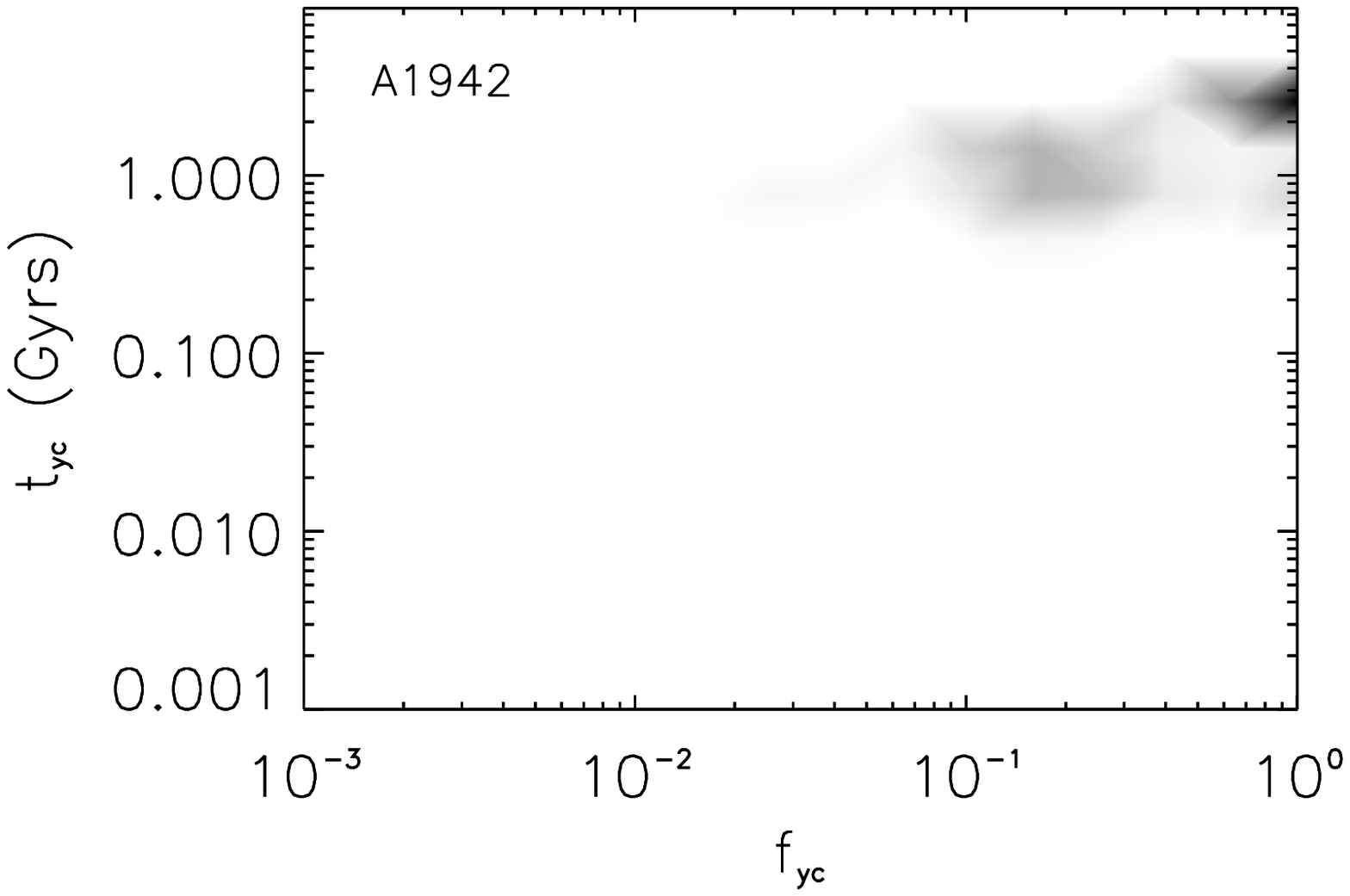}\\
   \includegraphics[width=3.5in]{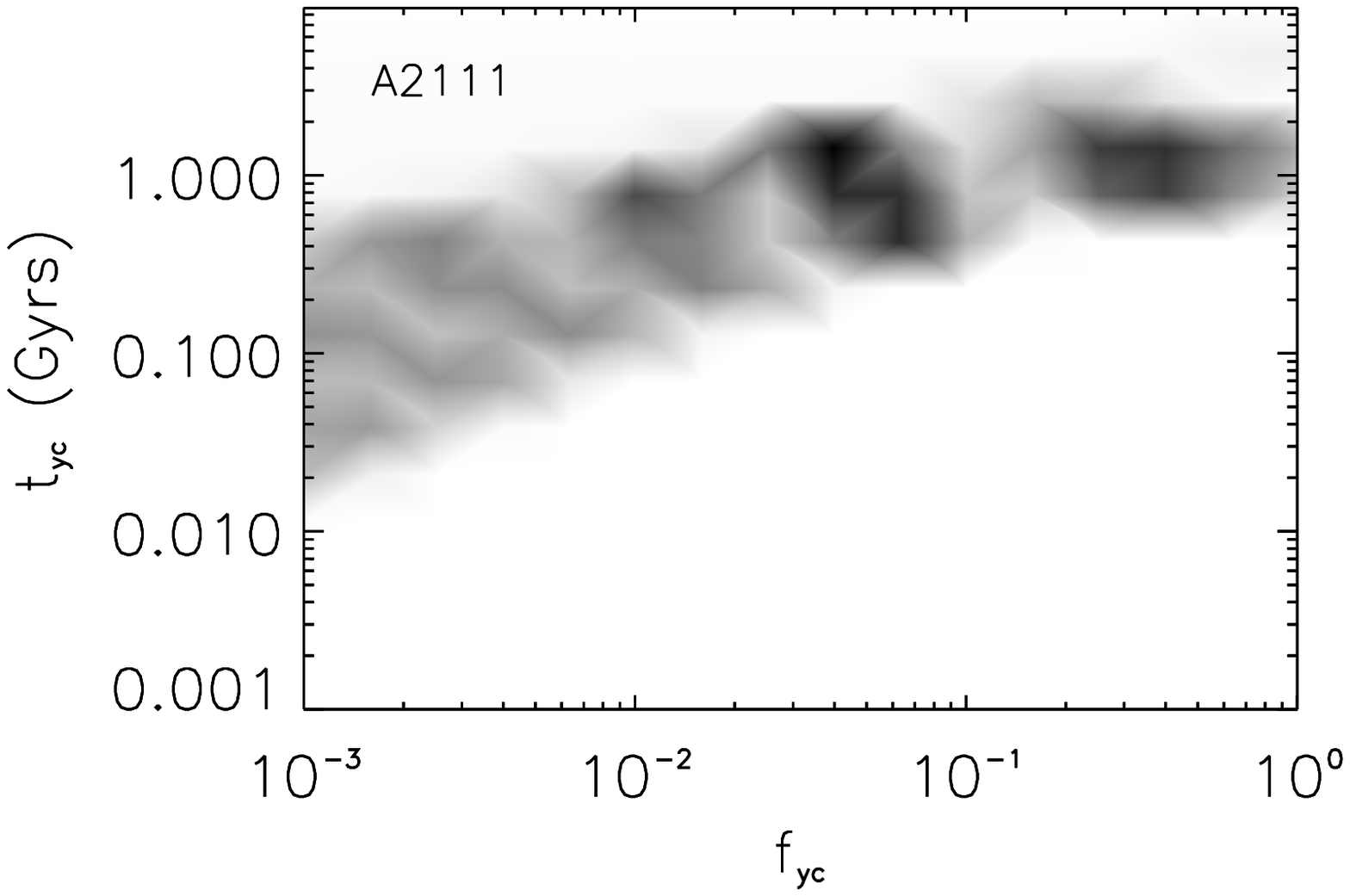}\\
   \includegraphics[width=3.5in]{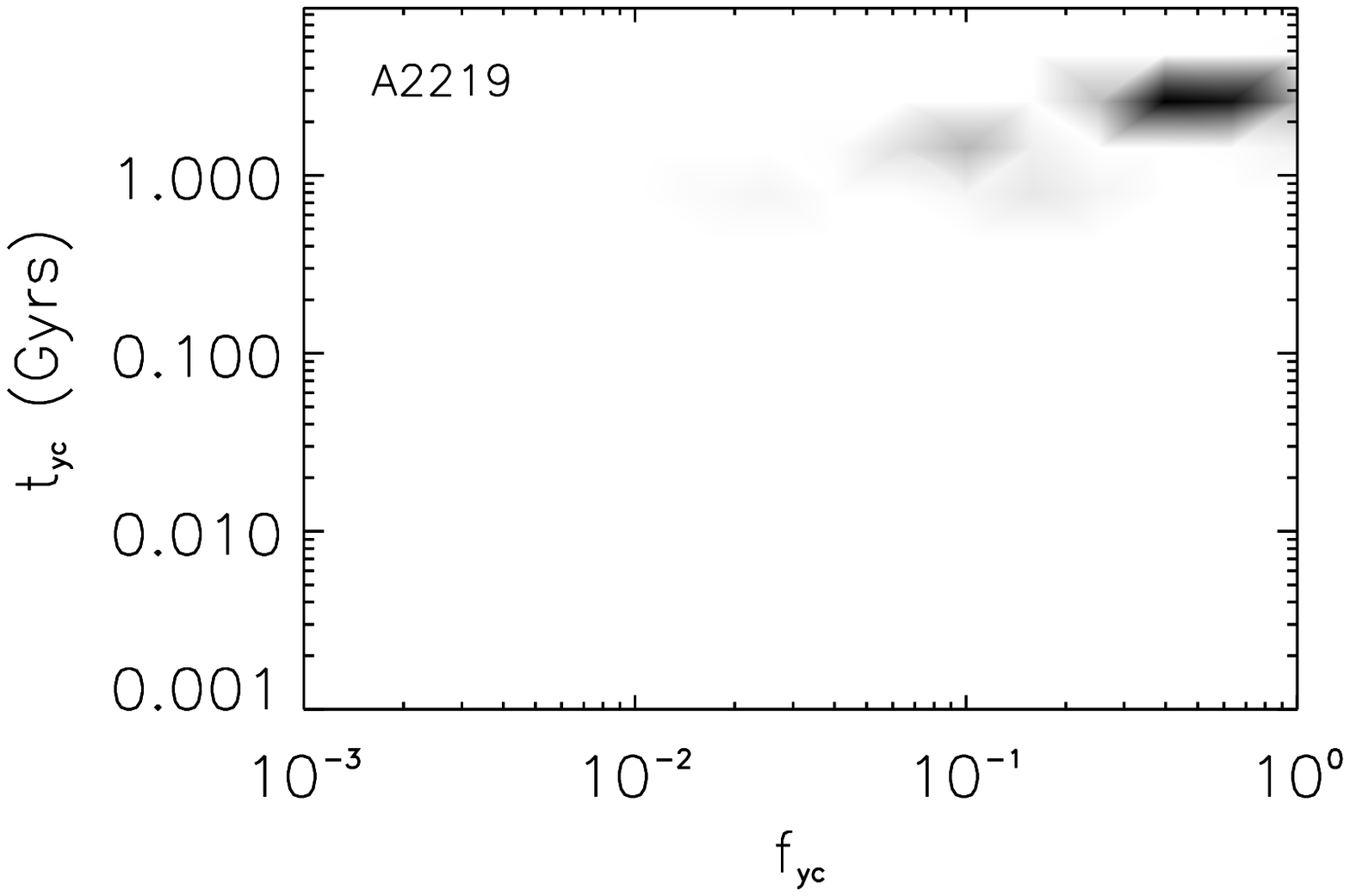}
   \end{array}$
   \caption{Probability maps for $t_{yc}$ and $f_{yc}$ for the UV-red
   BCGs. Recall that $t_{yc}$ and $f_{yc}$ represent the age and
   mass fraction of the recent star formation in the galaxy.
   From top to bottom the BCGs shown are: A1942, A2111 and A2219.
   We find that the `young component' (`yc') in each of these galaxies is composed of
   intermediate age populations with ages greater than 1 Gyr. Note that none of the galaxies
   are consistent with purely old ($>3$ Gyr old) populations. This is consistent with the fact
   that all the UV-red BCGs lie at least 1 mag away from the red sequence
   positions at their given redshifts.}
   \label{fig:prob-red}
\end{figure}

Two results are immediately apparent on inspection of  Figs.~\ref{fig:colours}--\ref{fig:prob-blue}. 
First,  the BCGs that do not exhibit blue cores (i.e.~nos. 2, 3, 7) are much closer to the NUV 
color of their respective red sequence positions: $d(NUV-r)<1.5$.  In other words, the red optical color corresponds
to red NUV color.  This is  not necessarily surprising but it does provide us with a baseline. What is 
interesting, however,  is that all of these  UV-red BCGs are somewhat displaced by from their  respective red sequence 
positions, suggesting that the star formation 
history of even these  UV-red BCGs is inconsistent with their stellar content having formed 
entirely at high redshift  ($z\sim3$) and that they likely  experienced low-level star formation 
events within the past  $\sim 2$ Gyrs.   This is the `time  window' during which the UV signal 
remains detectable (see top panel of Fig. 7 in Kaviraj et al. 2007a).  This conjecture is borne out by
the fact that the ($t_{yc}$, $f_{yc}$) map for the red-core BCGs (Figure \ref{fig:prob-red}) is not
concentrated exclusively at very old ages.  In the case of  A2111 BCG, for example, the recent star 
formation is consistent with intermediate age populations (since $t_{yc}$ is around 1 Gyr and 
$f_{yc}$ is around a few percent).

Given our small sample, we 
cannot reject the possibility that the $d(NUV-r)$ for our red core BCGs is likely biased 
high due to the requirement that they must be detected in the UV before entering the sample.  
Based on this, we would expect that any missing red core systems would likely lie even closer to the 
red sequence than the ones shown here, further emphasizing the gap in UV color between 
the optically blue core BCGs and their red core counterparts.  On the other hand, the offsets exhibited
by our red-core BCGs appears to be comparable to those of the local early-type population as a whole
and suggest that low-level star formation events are not uncommon or restricted to BCGs only.  
A careful analysis of the UV light in a large sample of elliptical galaxies indicates that 
luminous elliptical galaxies form up to  10-15 \% of their stellar mass after $z\sim1$, although for 
the bulk of the population, the typical young stellar mass fraction is much smaller, around a 
few percent (Kaviraj et al. 2007a; 2008; see also Kaviraj 2008 for a review of these recent results).   

Second and perhaps most interestingly, Figs.~\ref{fig:colours}--\ref{fig:prob-blue} reveal an unambiguous  
\emph{one-to-one correspondence between
UV `blueness' and a blue core in the optical image} in our BCG sample.  Every BCG that is far from the red sequence
position (i.e. has $d(NUV-r)>2.5$), which is rarely seen in the regular ellipticals, is also a blue core system, and vice versa.
Given that we associate the excess UV flux with star formation, this one-to-one correspondence, combined with the fact 
that UV-red BCGs do not show anomalous colour profiles,  bolsters our earlier assertions (as well as those of 
Bildfell et al. 2008) that the blue-core BCGs show the features that they do because they host active star formation in their 
central regions.  This association is further strengthened by independent confirmation of ongoing star formation based on optical
and infra-red spectroscopic work in 
three of our four blue-core systems.  These three galaxies and the measured rate at which they are forming stars in their
central regions are as follows:  A2390 is estimated to be forming stars at $\sim 5\;M_\odot\;$yr$^{-1}$,  CL0910 at
$\sim 40\;M_\odot\;$yr$^{-1}$ and A1835 at $\sim 120\;M_\odot\;$yr$^{-1}$ (see Table 3 in Bildfell et al.
and references therein).   

Setting aside the results for MS1512 for reasons noted at the beginning of this Section, 
Figure \ref{fig:prob-blue} indicates that the recent star formation in the blue-core BCGs typically has an age 
less than 200 Myrs and contributes mass fractions of less than a percent.  { Using the values of the most likely $t_{yc}$ 
and $f_{yc}$ inferred for the young stellar component in our blue-core BCGs,  we can estimate an 
upper limit for the SFR in the last $\sim$  0.2 Gyr for the UV-blue
ellipticals by assuming that a mass of stars $f_{yc}\times M_{*,BCG}$ is assembled
over a time $t_{yc}$.   The rates span the range $\sim 20-100 M_{\odot}/yr$.  These rates are broadly
consistent with the published star formation rates determined by other means though we note that the 
agreement is much better for systems with high star formation rates and that our derived rates tend to be higher
than the rates derived from spectroscopy for systems with low star formation rates. That our results tend
to be slightly higher is perhaps not a surprise especially when one allows for
the fact that we have assumed that all of the observed UV flux from the blue-core BCGs 
comes from the stellar component whereas this flux may include some (small) contribution from the AGN.   One 
could also argue that even in the absence of any AGN contamination, our estimated star formation
rates are upper limits because of the way we have chosen to model this process as an instantaneous burst, as opposed to a 
prolonged event that extends towards the present.   While true in general, we do not expect this to be an issue here 
because the age of the young stellar population in our blue-core galaxies is  comparable to the typical lifetimes of O 
and B stars.  In such cases, the extended and instantaneous models should both give similar results. }


\begin{figure*}
\begin{minipage}{172mm}
$\begin{array}{cc}
   \includegraphics[width=3.5in]{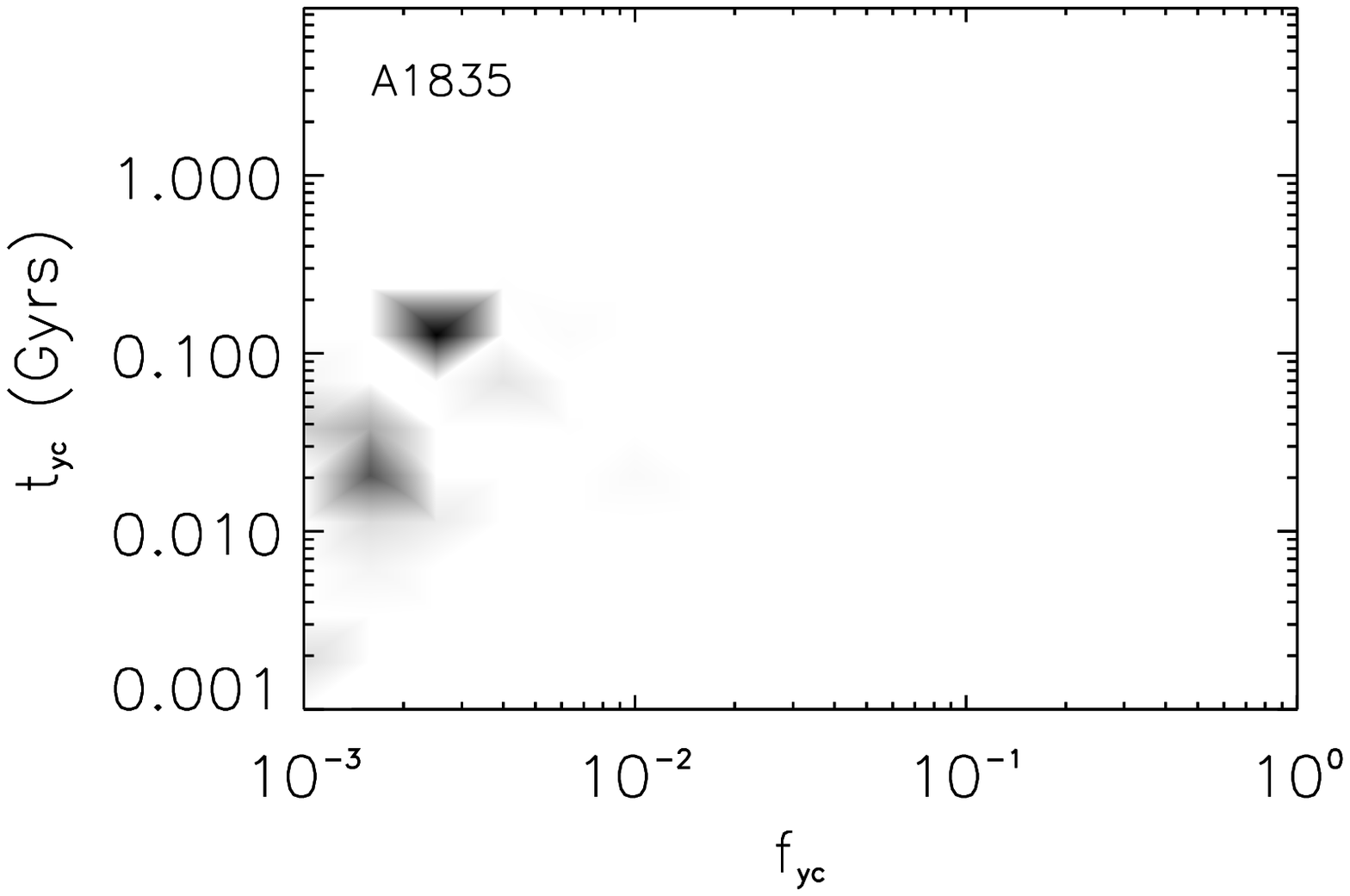}  & \includegraphics[width=3.5in]{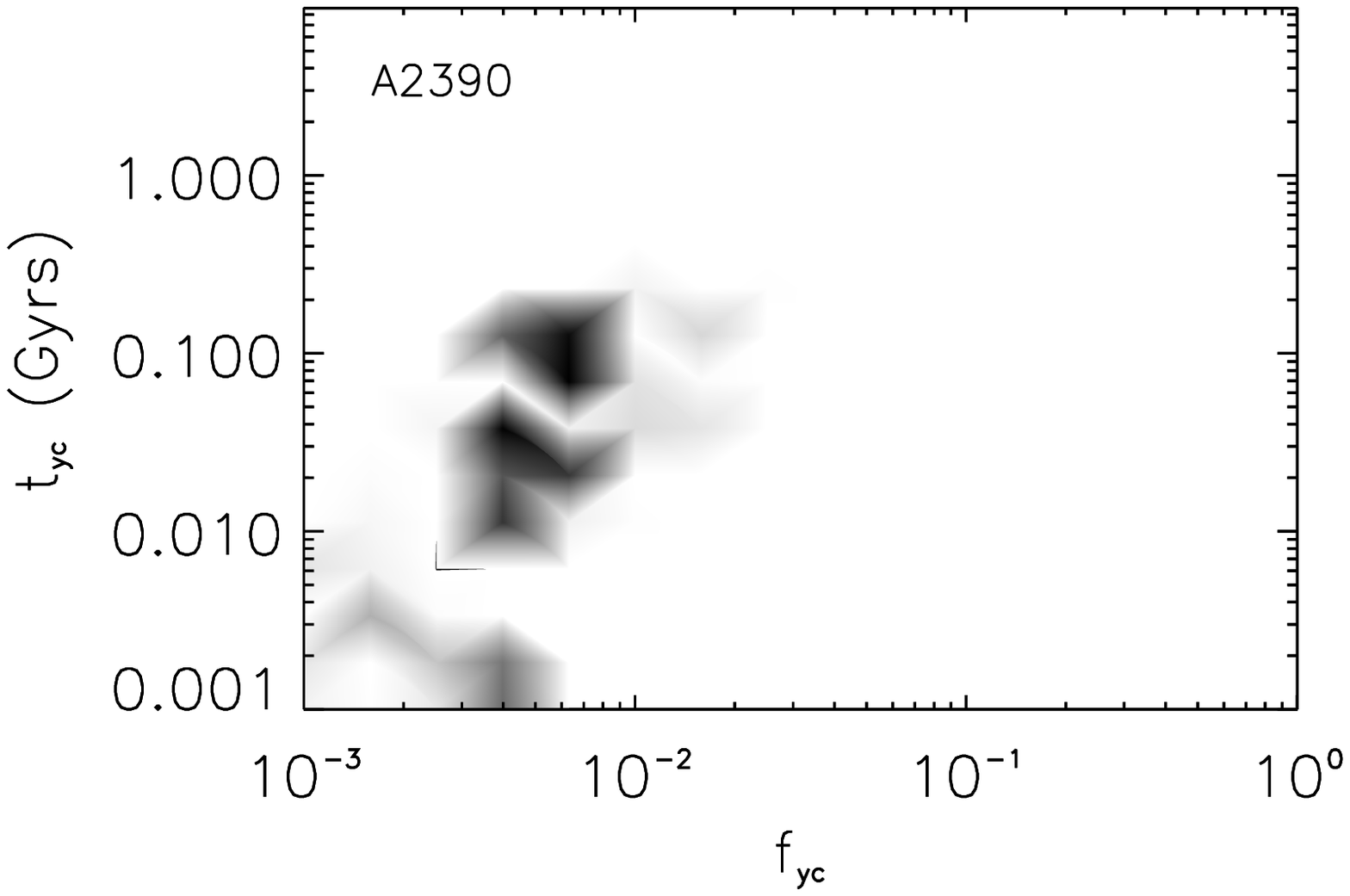}\\
   \includegraphics[width=3.5in]{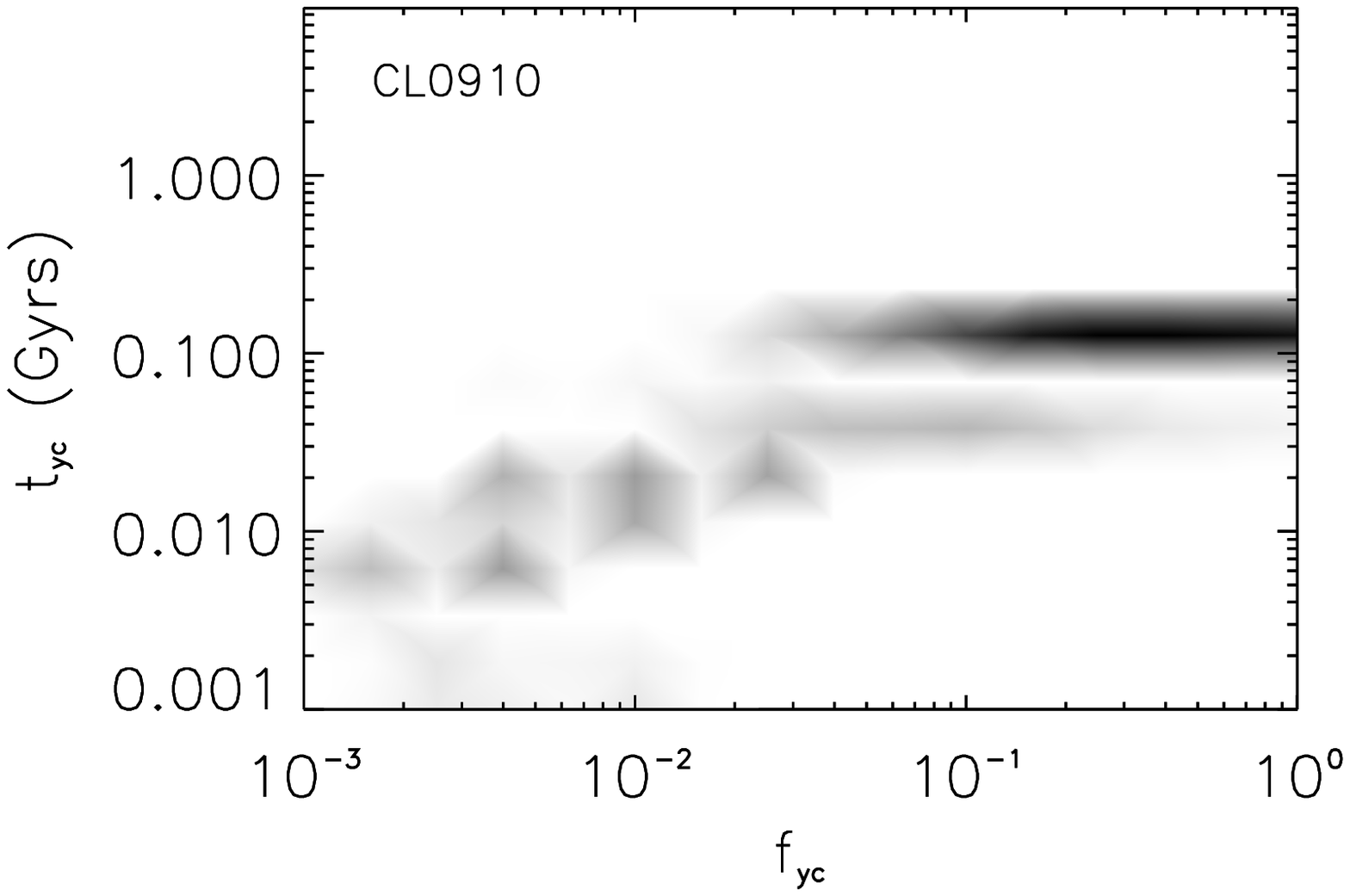} & \includegraphics[width=3.5in]{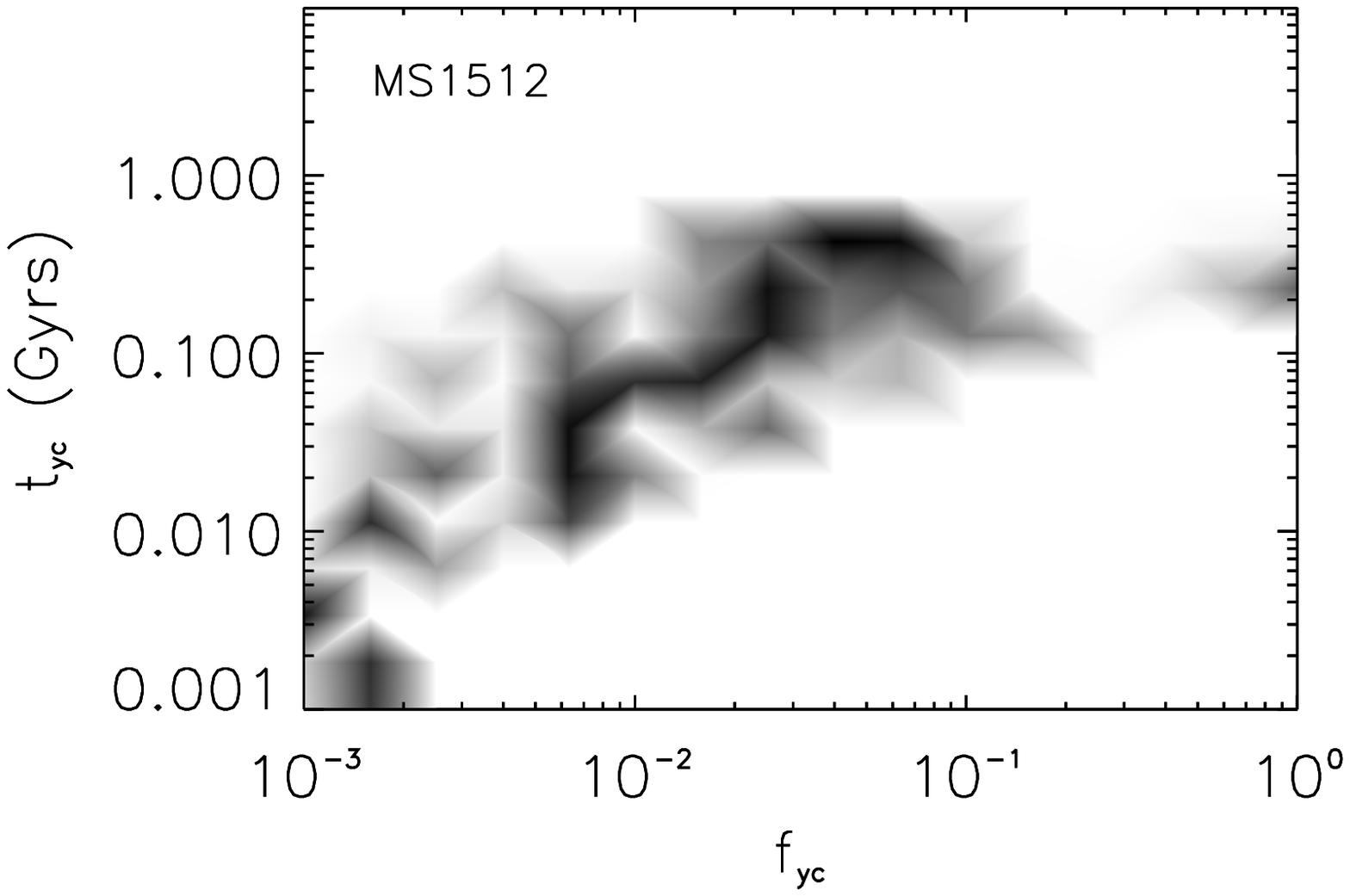}
   \end{array}$
   \caption{Probability maps for $t_{yc}$ and $f_{yc}$ for the UV-blue
   BCGs. Recall that $t_{yc}$ and $f_{yc}$ represent the age and
   mass fraction of the recent star formation in the galaxy. The
   recent star formation in the blue BCGs typically has an age less
   than 0.2 Gyrs and contributes mass fractions of less than a
   percent.}
   \label{fig:prob-blue}
\end{minipage}
\end{figure*}


\subsection{Plausible mechanisms for fueling the recent star formation in UV-blue BCGs}
\label{sec41}

Several of our blue-core BCGs, including A1835 and A2390, show evidence 
for significant accumulation of dust and molecular gas in their central regions (Edge et al. 2002, Egami et al. 2006).  
This reservoir is believed to provide the fuel for the ongoing star formation but where did this gas come from?  
Three obvious sources are: (1) gaseous flows associated with the cooling of the intracluster medium, (2) gas deposition during mergers,
and (3) gas ejected from the stars that comprise the BCG (Menanteau et al. 2001a; 2001b).   

Of these, Bildfell et al (2008) strongly favor the first on the strength of the correlations between the presence (or absence) of star formation 
in the BCGs and the global X-ray properties for the host clusters.   Bildfell et al. (2008) showed that the blue-core BCGs are exclusively located
in clusters that define the high-luminosity edge of the scatter in the $L_x$-$T_x$ plane.   In other words, 
at a fixed temperature, only clusters with the highest $L_x$ display observable signatures of active star formation.   

This region of the  $L_x$-$T_x$ diagram 
is known to be populated predominantly by cool-core clusters.   In the absence of a heating mechanism that can fully compensate for the efficient loss of thermal energy
through radiative losses, the cooling gas would be expected to flow inward towards the center of the potential, giving rise to cooling flows.  Cooling
flows provide a natural way to account for the reservoir of molecular gas in detected in some of the BCGs.   More recently, the link  between the central 
cooling timescale of the intracluster medium and star formation was further strengthened by the results of  Cavagnolo et al (2008) and Rafferty et al. (2008), 
who used Chandra observations to determine the central gas densities and temperatures for a large number of clusters and found that star formation 
typically only occurs in those clusters with central  cooling times $t_{\rm c} < 1$ Gyr.}  Interestingly, a detailed analysis of the individual BCG-cluster
systems in their CCCP sample led Bildfell to discover that while being embedded in a cool core cluster is a necessary condition for recent star 
formation in a BCG, it is not a sufficient condition.  The BCG must also lie at the centre of the cluster potential.  It is the combination of these
two trends that led Bildfell et al. (2008) to support the cooling flow hypothesis.   

The fact that all of our blue-core systems show evidence of a young stellar population ($<$ 200 Myr) suggests that either we have
caught these systems within 200 Myrs of their most recent bout of star formation, or that these systems have been forming stars for most of 
their lifetime.   We find the latter to be much more plausible especially since \emph{all} the BCGs in the CCCP sample that are hosted by clusters 
with high central X-ray surface brightness and are situated within $\sim10$h$^{-1}\;$kpc of the 
center of the host cluster potential (identified as the peak of the  X-ray surface brightness distribution) are blue-core and hence, actively star forming 
systems.\footnote{These systems comprise $\sim 20\%$ of the full CCCP systems.}     

If the star formation in the centrally located BCGs in cool core clusters is indeed fueled by cooling flows, then the above-mentioned
results also provide insights into the nature of cooling flows.   In the absence of ongoing heating, the short central cooling 
times of observed cool core clusters imply that radiative cooling would rapidly establish extreme centrally-peaked gas density profiles and 
sharply declining  central temperature profiles, with gas temperatures tending towards zero at the cluster center.   However, Chandra and XMM-Newton
X-ray  observations have convincingly demonstrated that present-day cool core clusters are not in the state just described --- their central cores are 
warmer than expected (e.g., Peterson et al.\ 2003).   This has led to claims in the literature that the AGNs in the 
centrally located BCGs are injecting sufficient energy into the intracluster medium to offset radiative losses and prevent the formation of 
cooling flows.   

We do not dispute that the intracluster gas is being heated; rather, based on our results, we assert that once the cooling flow is established, it 
is essentially ``on" for most of the time and that heating only acts to temper the cooling flow, not prevent them.  This is contrary to strong 
feedback models of De Lucia et al. (2006).
We cannot preclude the possibility of temporary disruption by mergers or AGN activity, only that  if disrupted, they reform
quickly.   That is, in models where AGN feedback actually manages to heat the cluster gas sufficiently so as to  halt the flow of cold gas towards 
the cluster center, such a phase must have a relatively short lifetime ($< 200$ Myrs), after which the flow and the associated star formation resumes.
Otherwise,  we would expect to see a fraction of the centrally located BCGs in cool core systems 
showing ``E+A"-like characteristics (i.e. the latest generation of stars with $t_{yc}\sim$ 1 Gyr).    Admittedly, our sample is small and 
we are mindful of that; however, the results for our blue-core systems stand in stark contrast to those of the BGCs in non-cool core clusters, in which the 
youngest stellar populations have ages $>$1 Gyr.     

Regarding the other two potential explanations for the cold gas and star formation, (Kaviraj et al. 2007a) find the recycling of 
the internal reservoir of gas built up from mass loss from the extent stellar population (Menanteau et al. 2001a; 2001b) is not 
efficient enough to render the  ellipticals UV-blue.   Moreover, if this was the explanation, one would also expect the BCGs
in non-cool core clusters to show evidence of current star formation, at variance with the findings of Bildfell et al. (2008) and 
Rafferty et al. (2008).   Additionally, one would expect that central star formation fueled by recycled gas would tend to preserve
and possibly even reinforce the radial metallicity gradients that is typically found in BCGs\footnote{A decrease of the metal
content in the stars of -0.3 dex per decade in radius is typically
observed with mild (if any) correlation with galactic mass in ellipticals (e.g.
Carollo et al. 1993) and BCGs (e.g. Fisher et al. 1995, Brough et al. 2007).} (see Pipino et al. 2006,
Pipino et al. 2008).   Cardiel et al. (1998), however,  found that in BCGs featuring emission-lines (due to recent star formation),
the gradients in the spectral indices are flat or even positive inside the emission-line regions. 
Outside the emission-line regions, and in cooling flow galaxies without emission lines, 
gradients are negative and consistent with those measured in giant elliptical galaxies
(Fisher et al., 1995). { At the same time, a residual star formation can lower the central [$\alpha$/Fe] abundance
ratio (e.g. Pipino \& Matteucci, 2006) and make blue BCGs differ from massive ellipticals.}
Larger samples, as the one by Loubser et al. (2008) will shed more light on the details of the 
chemical evolution in BCGs compared to normal ellipticals, providing additional clues into the source of the gas that is fueling the
observed star formation and to the integral star formation histories.

The accretion of gas-rich galaxies by the central BCG is another mechanism by which the latter can acquire its reservoir of gas.  
In fact, this mechanism may seem to be plausible for BCGs because they are likely to have more  companions 
than the average elliptical galaxy in the field and because dynamical friction within the cluster halo is likely to facilitate the orbital
decay of these galaxies.    Kaviraj et al. (2007c) find that galaxy mergers of mass ratios less than 1:4 and where the 
cold gas amounts to 20\% of the merging satellite mass, produce good agreement with the observed UV colors of regular 
ellipticals.  There are, however, several
problems with this particular explanation, of which the two most important are:   (a) The galaxies orbiting in or falling through the inner
cluster environments are typically gas poor; it is believed that they have either lost their gas to the intracluster medium via 
ram pressure stripping, or it was consumed by star formation and not replenished because of the nature of the local environment.
(b) This mechanism for delivery of cold gas should, to first order, be equally applicable to  
centrally located BCGs 
in both cool core and non-cool core clusters. Consequently, we would expect to see blue-core galaxies in both types of clusters.  We 
don't.   Additionally, there does not appear to be any outright correlation between presence or absence of a blue-core and the local
density of galaxies in the neighborhood of the BCG.  
For instance, MS1455\footnote{Not analysed here, but present
in the Bildfell et al sample} has very few satellites but exhibits a blue core while  
Abell 2537$^6$ has a red core but many satellites.

Of the three mechanisms discussed above, the \emph{cooling flow} explanation seems to be the most promising primary mechanism for
fueling the observed star formation.

{ Finally, the confirmation of recent star formation in early-type galaxies
calls for a dedicated survey to detect Type II supernovae (SNII)
in these systems. From a theoretical point of view, in fact, the current observational upper
limit in the central parts of galaxy clusters (0.027
SNuM\footnote{SNuM = SN explosion per century per
$10^{10}M_{\odot}$}, Mannucci et al. 2008) poses a strong constraint
on the recent star formation even if this value is slightly higher
than for the ellipticals as a whole.
In principle, we should be able to observe SNII events
in blue core BCGs. 
However, the typical brightness of SNII, combined
with the fact that the recent star formation takes place in
the bright centres of the galaxies, implies that the probability
of finding SNII in such galaxies using the current generation of
surveys is probably low. In fact, according to Mannucci et al.
(2007), 5-10\% of the local SNII remain out of reach of current
supernova surveys. The fraction of missing events rises sharply
towards $z=1$, when about 30\% of the SNII might be undetected.
Nevertheless, some good candidates do exist, such as the BCG in
A1835, which have very high star formation rates across an
extended star forming region, enhancing our chances of finding
SNII. 
Furthermore, the SNIa rate seems to be slightly
higher in radio-loud galaxies, indicating that recent star
formation may have boosted the \emph{prompt} channel of SNIa
progenitors (for details see Della Valle et al. 2005, Mannucci et
al. 2006). }

\section{Conclusions}
\label{conclusions} 

In a recent study, Bildfell et al. (2008) showed
that the majority of Brightest Cluster Galaxies (BCGs) in the
Canadian Cluster Comparison Project (CCCP) have shallow optical
colour profiles that become bluer with increasing radius. However,
a substantial minority of the BCGs in the  sample (25\%) deviates from this
simple behaviour, exhibiting blue optical cores instead. In this
study, we have presented a joint analysis of the near ultra-violet (NUV) data
from the GALEX mission and optical colour profiles for a  sample of
7 BCGs from the Bildfell et al. sample.

We find that every BCG that has a blue  NUV colour also
shows a blue-core in its optical colour profile. Conversely, BCGs
that lack blue cores and show monotonic colour gradient are red in the UV.
Although the sample studied here is small, we demonstrate,
\emph{for the first time}, that a one-to-one correspondence between
blue cores and  UV-enhancement is a clear indicator of recent star formation.
The UV-enhancement is not due to  old evolved stellar populations such as 
horizontal branch stars.  While we cannot outrightly rule out a contribution to the 
observed UV-flux from an obscured central
AGN in the BCG, a number of factors including the physical size of the blue
cores and the fact that obscured AGNs are expected to contribute to the optical signal at
a few percent level lead us to assert that AGNs are not the principal agents
of the optical and NUV observations.    Our assertions are in agreement with 
other independent indicators of star formation in 
BCGs based on optical and infra-red observations (e.g. Cardiel et al. 1998, Crawford et al. 1999, Edge 2001, Goto 2005,
McNamara et al. 2006, Wilman et al. 2006, Bildfell et al. 2008, Cavagnolo et al. 2008, O'Dea et al. 2008, Rafferty et al. 2008).

The young stellar component in all of our blue core BCGs typically has an age 
less than 200 Myrs.   The presence of ongoing star formation strongly implies that 
these systems have been forming stars for most of their life time, albeit at a low rate.  Given 
that  \emph{all} the BCGs in the CCCP sample that are hosted by clusters 
with high central X-ray surface brightness and situated within $\sim10$h$^{-1}\;$kpc of the 
center of the host cluster potential  are blue-core, we must accept that all such systems must
be steadily forming stars over a cosmological time.

We discuss several possible sources of gas for feeding the recent star formation in the blue core BCGs: (a) cooling flows 
(Bildfell et al. 2007); (b) recycling of stellar ejecta (Menanteau et al. 2001); (c) mergers with gas-rich galaxies
(Kaviraj et al. 2007c).  However, an analysis of each of these possibilities combined with unambiguous linkages between 
the color of the BCG core, the location of the BCG within the host cluster, as well as the excess X-ray luminosity (Bildfell et al. 2008) of
the cluster and the short central cooling time of the intracluster gas (Rafferty et al. 2008) leads us to conclude that the 
cooling flow is the primary mechanism for fueling the recent star formation BCGs.  In this regard, BCGs seem to behave
differently from the rest of elliptical galaxies.  This also means that heating from the AGN largely moderates the cooling
flow, as opposed to preventing its formation altogether like in the strong feedback model of De Lucia et al. (2006).

Finally, we note that while we have demonstrated a direct correspondence between the
presence of a blue optical core and a blue UV colour in a small
sample of BCGs, the result needs to be confirmed through an analysis of a much
large sample of elliptical galaxies, drawn from a
wide range in luminosity and environments.   Among the interesting issues to consider
is whether BCGs in group environments because similarly to our cluster BCGs.
Future work will focus on repeating the analysis presented here on GALEX-detected
early-type galaxies that have SDSS images (from which colour
profiles can be extracted) to test the robustness of the
preliminary results presented in this study.


\section*{Acknowledgments}
The authors thank the referee for his careful reading and his insightful comments.
Enlightening discussions with F.Mannucci, R.Maiolino, I. McCarthy and B. McNamara are acknowledged.
AP acknowledges partial support from NSF grant AST-0649899.
SK acknowledges research support through a Leverhulme Early-Career Fellowship, a BIPAC
Fellowship and a Research Fellowship from Worcester College, Oxford.   AB and HH acknowledge support 
from NSERC through the Discovery Grant program. 

GALEX (Galaxy Evolution Explorer) is a NASA Small Explorer,
launched in April 2003, developed in cooperation with the Centre
National d'Etudes Spatiales of France and the Korean Ministry of
Science and Technology.

Funding for the SDSS and SDSS-II has been provided by the Alfred
P. Sloan Foundation, the Participating Institutions, the National
Science Foundation, the U.S. Department of Energy, the National
Aeronautics and Space Administration, the Japanese Monbukagakusho,
the Max Planck Society, and the Higher Education Funding Council
for England. The SDSS Web Site is http://www.sdss.org/.

The SDSS is managed by the Astrophysical Research Consortium for
the Participating Institutions. The Participating Institutions are
the American Museum of Natural History, Astrophysical Institute
Potsdam, University of Basel, University of Cambridge, Case
Western Reserve University, University of Chicago, Drexel
University, Fermilab, the Institute for Advanced Study, the Japan
Participation Group, Johns Hopkins University, the Joint Institute
for Nuclear Astrophysics, the Kavli Institute for Particle
Astrophysics and Cosmology, the Korean Scientist Group, the
Chinese Academy of Sciences (LAMOST), Los Alamos National
Laboratory, the Max-Planck-Institute for Astronomy (MPIA), the
Max-Planck-Institute for Astrophysics (MPA), New Mexico State
University, Ohio State University, University of Pittsburgh,
University of Portsmouth, Princeton University, the United States
Naval Observatory, and the University of Washington.


\end{document}